\documentclass[journal, 12pt]{IEEEtran}
\usepackage{xcolor, soul,framed} 

\usepackage[pdftex]{graphicx}
\graphicspath{{../pdf/}{../jpeg/}}
\DeclareGraphicsExtensions{.pdf,.jpeg,.png}

\usepackage[cmex10]{amsmath}
\usepackage{array}
\usepackage{mdwmath}
\usepackage{mdwtab}
\usepackage{eqparbox}
\usepackage{url}

\usepackage[labelsep=period]{caption}
\usepackage{float}

\begin{document}
\bstctlcite{IEEEexample:BSTcontrol}
    \title{Towards Intelligent RAN Slicing for B5G: Opportunities and Challenges}
  \author{EmadElDin~Mazied, 
      Lingjia~Liu, 
      and~Scott Midkiff


  
  }



\maketitle

\begin{abstract}
To meet the diverse demands for wireless communication, fifth-generation (5G) networks and beyond (B5G) embrace the concept of network slicing by forging virtual instances (slices) of its physical infrastructure. While network slicing constitutes dynamic allocation of core network and radio access network (RAN) resources, this article emphasizes RAN slicing (RAN-S) design. Forming on-demand RAN-S that can be flexibly (re)-configured while ensuring slice isolation is challenging. A variety of machine learning (ML) techniques have been recently introduced for traffic forecasting and classification, resource usage prediction, admission control, scheduling, and dynamic resource allocation in RAN-S. Albeit these approaches grant opportunities towards intelligent RAN-S design, they raise critical challenges that need to be examined. This article underlines the opportunities and the challenges of incorporating ML into RAN-S by reviewing the cutting-edge ML-based techniques for RAN-S. It also draws few directions for future research towards intelligent RAN-S (iRAN-S).
\end{abstract}


%
\IEEEpeerreviewmaketitle


\section*{Introduction}
Network slicing grants fifth-generation networks and beyond (B5G) an opportunity of provisioning diverse reconfigurable network environments to support distinct use-cases, which each has unique demands.
For example, telemedicine, smart homes, and road safety use-cases need to be supported by enhanced mobile broadband (eMBB) services, massive machine-type-communication (mMTC), and ultra-reliable and low latency communication (URLLC), respectively. 
Network slicing conceptualizes the B5G design by sharing network resources to form \textit{\textbf{virtual}} network instances (\textit{\textbf{slices}}) that are \textit{\textbf{isolated}} and \textit{\textbf{customized}} to satisfy \textit{\textbf{diverse demands}} for communication services within a specific period (\textit{\textbf{slice lifetime}}).
  
Contriving end-to-end network slicing considers dynamic management of B5G's core network (CN) and radio access network (RAN) resources. However, B5G's RAN includes more dynamic characteristics than its CN, i.e., RAN includes heterogeneous radio resources and functions. Therefore, we accentuate the RAN slicing design (RAN-S) in B5G.

Thanks to function disaggregation and resource virtualization principles that bring RAN-S to B5G. Function disaggregation reflects RAN's tendency to be programmable through software-defined radio (SDR) and software-defined network (SDN) paradigms. Resource virtualization renders virtualizing network functions, computing, and communication resources to provide a reconfigurable service architecture for B5G~\cite{CompInComm}.

\begin{figure}
\begin{center}
\includegraphics[width=1\columnwidth]{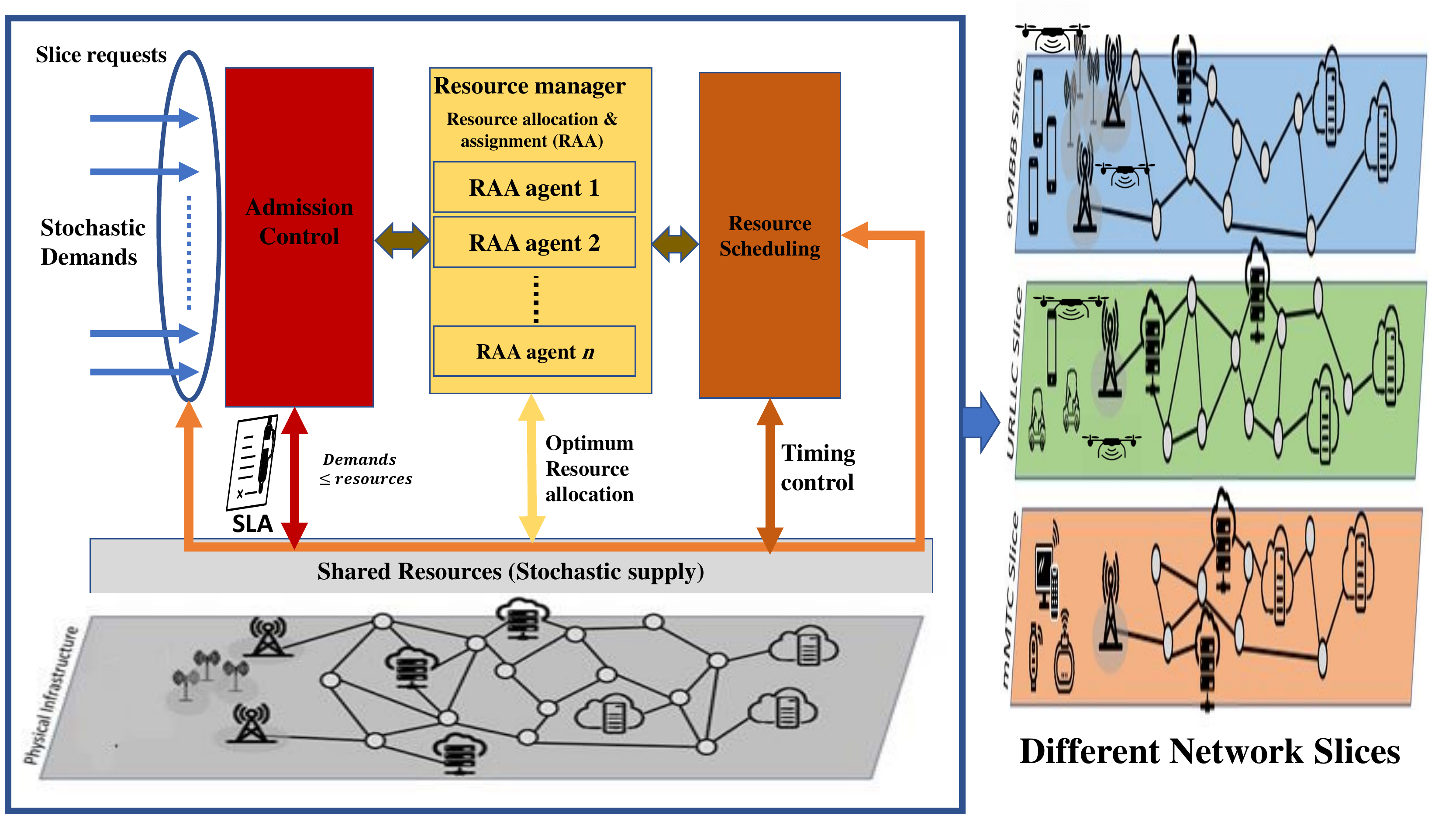}
\caption{Overview of network slicing design.}
\label{fig:NS_Arch}
\end{center}
\end{figure}

There are two players in RAN-S arena; infrastructure providers (InPs) and service providers (tenants). InPs strive for revenue maximization by magnifying resource utilization, while tenants seek to satisfy use-cases' service requirements. A broker, i.e., slice broker, manages the InPs-tenants relationship by settling service level agreements (SLAs) between them.

Figure~\ref{fig:NS_Arch} shows the RAN-S's slice broker that constitutes an admission control (AC) agent, a resource manager (RM), and a scheduling agent. The AC agent applies an admission policy regarding the tenants' requests for slices to assure SLA satisfaction. The RM, the broker's brain, dynamically (re)-allocates InP's resources to meet tenants' demands. The scheduling agent schedules allocated resources among admitted slices to guarantee fair resource sharing. Accordingly, RAN-S is deemed a \textit{\textbf{dynamic resource sharing}} application running a substrate of heterogeneous and distributed resources.

\textbf{\textit{``Dynamic''}} and \textbf{\textit{``stochastic''}} are interchangeable terms in the resource sharing jargon. Different stochastic tools, e.g., stochastic dynamic programming, were devised to optimize different design aspects in RAN-S; however, they provided near-optimal solutions. On the other hand, various machine learning (ML) techniques have recently demonstrated a better convergence to an optimal design than what stochastic tools provided for RAN-S~\cite{NS_meets_AI}. For example, to tackle the dynamics in demands, deep learning-based prediction methods were introduced in~\cite{DeepAI4MTDMobicom20, RsrcUsagePredictionMobiHoc19}. Besides, deep reinforcement learning (DRL) was adopted to develop AC policies~\cite{ML_Based_AC_5G_MarketOptim, RL_Flexible_AC}. Moreover, dynamic resource allocation and scheduling of computing and radio resources, where a high time-varying setting is present, were handled by DRL methods in~\cite{DL_based_Comp_RAN_Scheduling_Mobicom19, EdgeSliceWithDDRL, RL4DRA_RAN_Slicing, Intelligent_Rsrc_Schedu_5G_RAN_Slicing}. 

Nevertheless, the introduced ML schemes have some limitations that need to be scrutinized towards intelligent RAN-S (iRAN-S). For example, the prediction accuracy of traffic forecasting in~\cite{DeepAI4MTDMobicom20}, which would reflect dynamics in demands, was limited to infrastructure-based aggregate traffic. However, machine-type-communication-generated traffic could be considered for robust interpretations of the dynamics in demands. Later, we discuss additional challenges that pertain to the introduction of ML for RAN-S. We also envisage the incorporation of ML-based information leakage models~\cite{NN_autoencoder_Trans_Info_Forensics_Security, SCAUL} into RAN-S to tackle the challenge of slice isolation.

This article discusses various ML-based schemes that address the dynamics in demands and resources towards iRAN-S. In the following, we present the concept of network slicing. The article then explores the opportunities that ML would bestow for RAN-S and the raised challenges by reviewing recent advances in ML techniques for iRAN-S. Following that, the article draws a few directions for future research. Finally, we conclude the article.

\section*{The Concept of Network Slicing}
Diverse use-cases drive fifth-generation wireless networks and beyond (B5G) (e.g., health care, transportation, surveillance, etc.). However, meeting the demands for all use-cases with the same network settings is unattainable. For example, to provide unmanned aerial vehicles (UAV) with ultra-reliable and low latency communication (URLLC), most radio functions would be placed at the far-edge network. However, UAV-to-ground communication needs enhanced mobile broadband (eMBB) service for real-time video/image transmission, which requires higher computational power for video/image processing. Hence, to attain URLLC service, more computational resources would be provisioned at the far-edge; otherwise, transport and mobile edge computing (MEC) resources would be allocated for UAV's video/image processing. While the former approach would elevate operational expenditures (OPEX), the latter approach would violate URLLC. Therefore, to tackle this design tradeoff, network resources and functions need to be dynamically tailored to support UAVs with eMBB and URLLC during their mission time (i.e., use-case lifetime).

Tailoring B5G resources and functions resembles the dynamic allocation of computing resources to support concurrent execution of a sophisticated software system's threads, i.e., the concept of program slicing. In program slicing, software functions are divided (disaggregated) into multiple threads, and the computing resources are dynamically (re)-configured to provide virtual computing environments for parallel computation. Correspondingly, to meet B5G's use-cases demands, softwarization and virtualization are deemed the superior enabling-technologies for network slicing. 
As introduced earlier, B5G softwarization is its tendency to be programmable by software-defined radio (SDR) and software-defined network (SDN) technologies~\cite{CompInComm}.
Virtualization embodies resource sharing by leveraging B5G's programmability to manage its resources. It abstracts the demands, identifies the interface between demands and resources (i.e., hypervisor), specifies the methodology of assigning and accessing the allocated resources, and schedules time-slots for resource usage.  Simply put, virtualization is the software that drives resource controllers and schedulers by applying a resource-sharing algorithm. Virtualization includes physical virtualization and network function virtualization (NFV). The former represents the virtualization of physical resources, while the latter conveys different network functions' virtualization.

\begin{figure}
  \begin{center}
    \includegraphics[width=1\columnwidth]{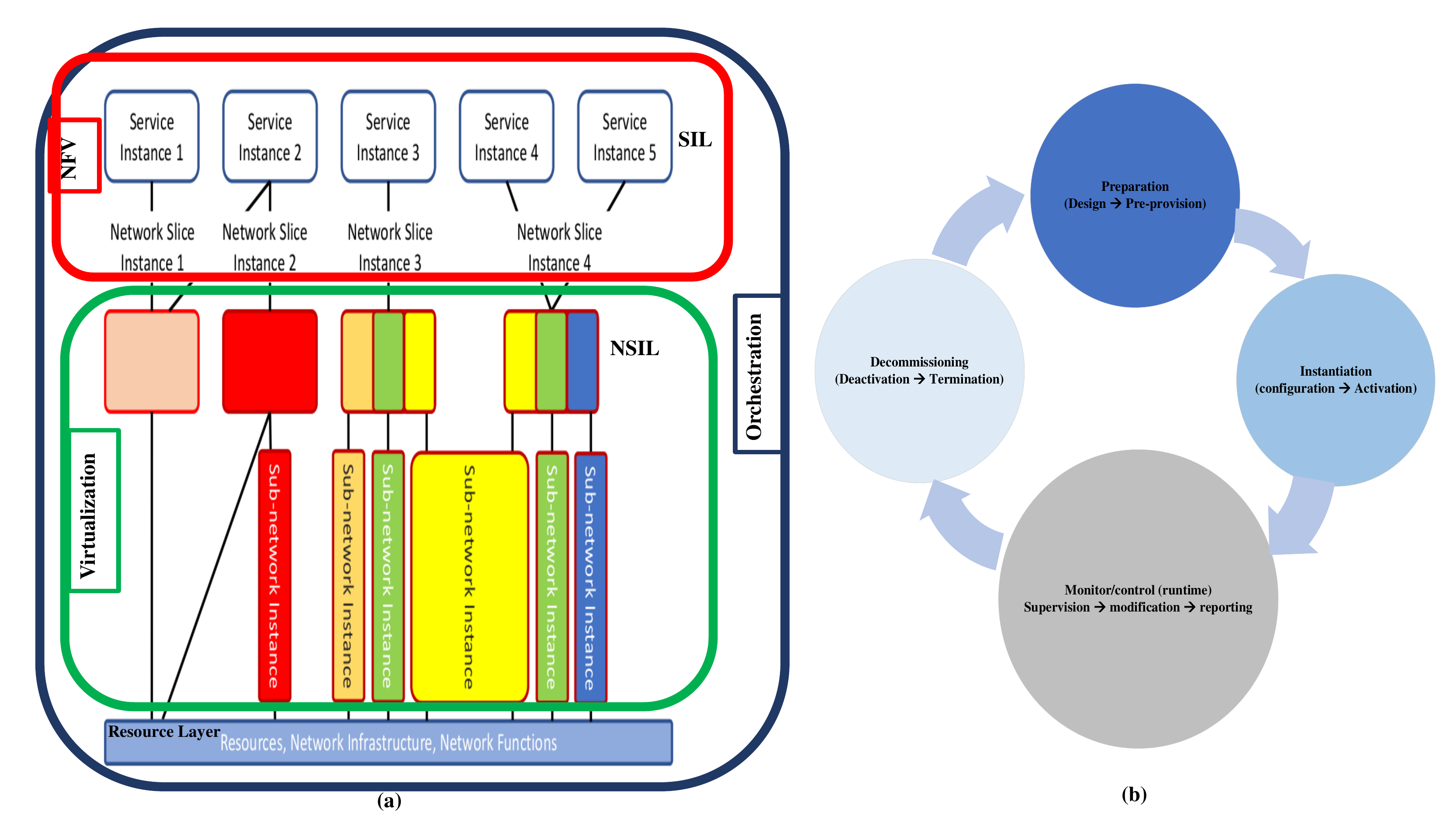}
 \caption{The concept of network slicing~\cite{CompInComm}: (a) Slicing architecture; (b) Slice life cycle.}
 \label{fig:NS_concept}
 \end{center}
\end{figure}

\textit{Dynamic} deployment of virtual networks (slices) would satisfy the demands for supporting a specific service category, e.g., massive machine-type-communication (mMTC), URLLC,  or eMBB. Each slice would be established and maintained with the most proper (re)configuration to meet each service category's requirements, provided that each slice can not affect the other slices (i.e., slice isolation). Thus, network slicing (re)defines how the network behaves and adapts its operation according to application requests (i.e., slice customization). Furthermore, different slices would serve the same end-users, e.g., eMBB-URLLC coexistence for UAV use-case.

Figure~\ref{fig:NS_concept} (a)~\cite{CompInComm} depicts the B5G slicing architecture. The resource layer abstracts the network infrastructure and hosts various network slicing sub-network instances (NSSIs). By virtualization tools, network slice instant layer (NSIL) forms network slice instants (NSIs), which each includes sub-network resource instants, i.e., NSSIs. Each NSSI comprises a set of physical network functions (PNFs) or virtual network functions (VNFs). NSI can share different NSSIs. Relying on NFV, the service instant layer (SIL) supports different services for end-users (tenants) by utilizing the underlying NSIs. Each service can utilize one or more NSIs. The network slice controller (NSC) orchestrates the procedures above through an open network operating system (ONOS). ONOS interfaces with each layer's functions to enable flexible formation and reconfiguration during the slice life cycle. 

Figure~\ref{fig:NS_concept} (b)~\cite{CompInComm} illustrates the process that NSIL follows to request, instantiate, monitor, and terminate a slice. The process begins with designing and preparation of a slice template. Following that, an instantiation request to forge, configure and activate the slice. Then, during slice operation, the slice performance is tracked and controlled to accomplish service level agreement (SLA). Finally, NSC terminates the slice when it is no longer needed, and the allocated resources can then be re-shared for other slice operations. Network slicing would require the deployment of multiple NSCs to subdue the system intricacy. Each NSC would manage a subset of functionalities for each layer and exchange information about slice operation during its life cycle with other NSCs.


\section*{Recent advances in Machine Learning Techniques for Radio Access Network Slicing Design}
As mentioned earlier, network slicing includes core network (CN) and radio access network (RAN) slicing. However, the discussion of CN slicing is out of this article's scope.

Recent advances in machine learning (ML) would promote its incorporation into RAN slicing design (RAN-S). Authors in~\cite{NS_meets_AI} showed how ML's performance would outperform the conventional stochastic tools in terms of better convergence to an optimal decision for RAN-S. Therefore, we diagnose the recent advances in ML techniques towards intelligent RAN-S (iRAN-S) design in the following. Mainly, we highlight the developed ML techniques for dynamic resource sharing problems by reviewing the state-of-the-art related work to dynamics in demands, admission control, and dynamics in resources. Then, we discuss the slice isolation problem in the light of recent ML-based information leakage techniques.

\subsection*{Dynamics in Demands}
\begin{figure}
 \begin{center}
 \includegraphics[width=1\columnwidth]{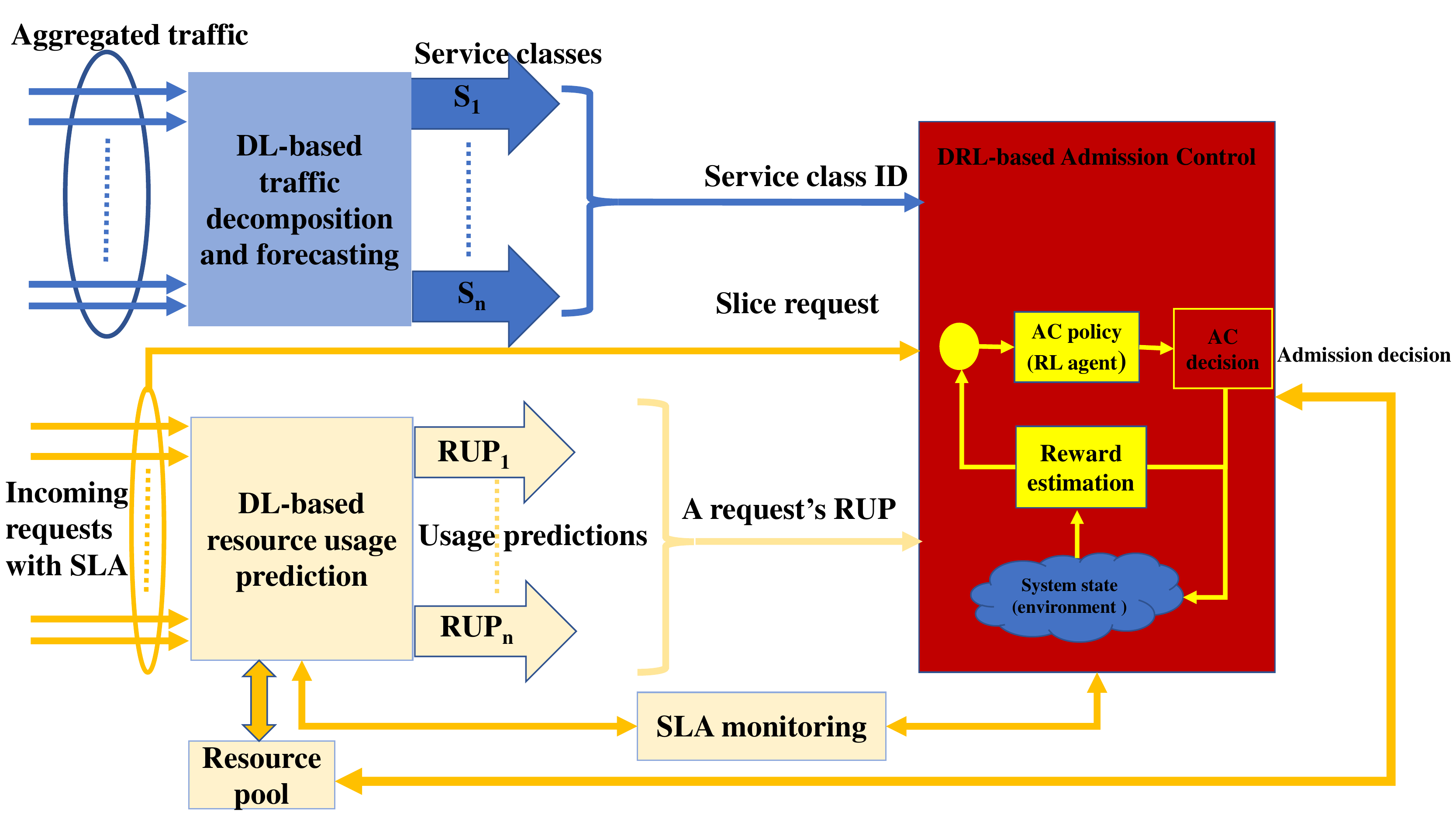}
 \caption{Dynamics in demands and admission control~\cite{DeepAI4MTDMobicom20, RsrcUsagePredictionMobiHoc19, ML_Based_AC_5G_MarketOptim, RL_Flexible_AC}.}
 \label{fig:dynamicsInDemandsandAC}
\end{center}
\end{figure}

Figure~\ref{fig:dynamicsInDemandsandAC} visualizes two concepts that are recently introduced to reflect dynamics in demands for iRAN-S~\cite{DeepAI4MTDMobicom20, RsrcUsagePredictionMobiHoc19}; traffic forecasting and resource usage prediction. 

In~\cite{DeepAI4MTDMobicom20}, mobile traffic from different RAN resources (e.g., radio towers and edge computing) was decomposed and then classified into various service classes. The authors adopted a deep neural network to predict and decompose different traffic classes from aggregate traffic by leveraging the spatiotemporal correlation in mobile traffic. 
Although this work would substantially contribute to iRAN-S, accuracy-delay trade-off and training data collection need to be examined. In this work, traffic decomposition relied on the Hungarian assignment algorithm for displacement optimization (i.e., $\mathcal{O}(n^3)$). Thus, when problem size ($n$) increases, processing delay for the traffic forecasting process increases, specifically, for the URLLC slice real-time operation. If we consider minimizing $n$, we ruin prediction accuracy. Additionally, the training data was collected from network infrastructure; however, considering ad hoc traffic, i.e., autonomous machines' traffic (vehicles, UAVs, etc.), is needed for better traffic prediction and classification.

Authors in~\cite{RsrcUsagePredictionMobiHoc19} introduced an indirect traffic forecasting module, i.e., resource usage prediction (RUP), that relied on predicting resource usage (RU) by a slice. The preceding measurements of (RU) interpreted the carried traffic that a slice has served. As shown in Figure~\ref{fig:dynamicsInDemandsandAC}, the predicted RU contributed to RA's admission control. It would also contribute to the allocation and scheduling of radio resources for subsequent slice requests as depicted in Figure~\ref{fig:NS_Arch}. RU was defined in terms of physical resource blocks (PRBs) and was predicted by learning from RU measurements' history. Nevertheless, in addition to PRBs, we should consider other RAN resources (i.e., computing, transport, storage, etc.) for enhanced RU forecasting. These resources are critically needed for different use-cases, i.e., UAVs need sufficient computational resources for real-time video transmission in addition to the URLLC service. 
As shown in Figure~\ref{fig:dynamicsInDemandsandAC}, we envisage that traffic decomposition and classification scheme in~\cite{DeepAI4MTDMobicom20} could complement RUP for a robust admission control design.

\subsection*{Admission Control}
Admission control (AC) has a significant impact on resource utilization, infrastructure provider's (InP's) revenue, and service requirements satisfaction of running slices. The AC decision to admit (reject) incoming slice request interprets resource allocation (release). Thus, an imprecise AC decision would result in either overbooking or underutilization of resources. Thus, keen AC decision-making is needed to tackle the trade-off between resource provisioning and SLA satisfaction.

Reinforcement learning (RL) has been introduced in~\cite{ML_Based_AC_5G_MarketOptim, RL_Flexible_AC} for AC design that would interact with the dynamics in RAN operation. While authors in~\cite{ML_Based_AC_5G_MarketOptim} deployed semi-Markovian Decision Process (SMDP) model for RL agent, authors in~\cite{RL_Flexible_AC} introduced a stochastic artificial neural network (S-ANN) to model the RL agent. Both models aimed to maximize long-term reward (InP's revenue).
The former worked on admitting (rejecting) new requests, but the latter tackled resource scalability for running slices by scaling up (down) the allocated resources. Both models trained RL agents by interaction with the environment to optimize the AC decision policy. However, authors in~\cite{ML_Based_AC_5G_MarketOptim} incorporated deep learning into the RL by introducing two feed-forward neural networks (FFNN), i.e., admission FFNN and rejection FFNN, to predict the long term reward function (InP's revenue) for each possible AC decision. Nonetheless, both AC schemes need to develop slice isolation criteria for AC design. Moreover, to develop the admission region in~\cite{ML_Based_AC_5G_MarketOptim}, the authors considered only two traffic classes (i.e., elastic and inelastic). Traffic was classified based on throughput metric, whereas examining other traffic types into the AC design is required. Furthermore, the validation topology assumed equally-spaced base-stations; however, different topological settings (i.e., macro, small, micro, pico-cell) need to be studied. Besides, Q-learning is deployed in~\cite{RL_Flexible_AC}, though it has an inherent limitation with large state-space problems.

\subsection*{Dynamics in Resources}

\begin{figure}
 \begin{center}
 \includegraphics[width=1\columnwidth]{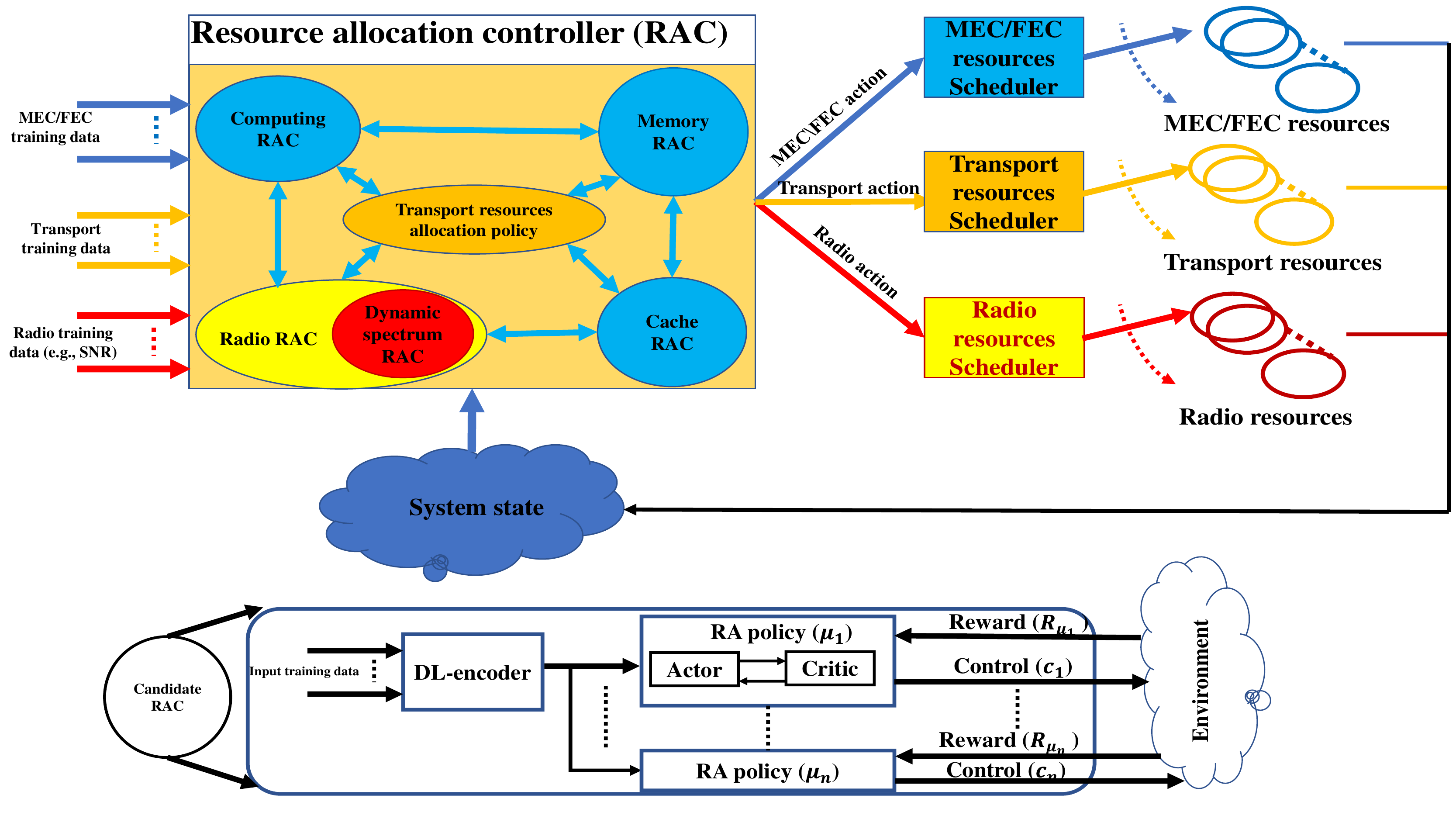}
 \caption{Dynamic resource allocation and scheduling~\cite{DL_based_Comp_RAN_Scheduling_Mobicom19, EdgeSliceWithDDRL, RL4DRA_RAN_Slicing, Intelligent_Rsrc_Schedu_5G_RAN_Slicing}.}
 \label{fig:DRAandSchedu}
\end{center}
\end{figure}

In the following, we explore the dynamics in resources in the context of dynamic resource allocation and scheduling.

Figure~\ref{fig:DRAandSchedu} fits together the recent advances in ML-based dynamic resource allocation (DRA) and scheduling for iRAN-S that were discussed in~\cite{DL_based_Comp_RAN_Scheduling_Mobicom19, EdgeSliceWithDDRL, RL4DRA_RAN_Slicing, Intelligent_Rsrc_Schedu_5G_RAN_Slicing}.

Authors in~\cite{DL_based_Comp_RAN_Scheduling_Mobicom19} developed deep reinforcement learning (DRL) and unsupervised deep learning (DL) schemes for DRA of RAN's radio and computing resources. They adopted contextual space and deep deterministic policy gradient (DDPG) algorithm to tackle the challenge of the wireless and computing environment huge state-space and the continuous action-space of the control environment, respectively. Distinct from conventional DRL, which defines state-space of all possible system's states, the introduced DRL-based resource manager (DRL-RM) utilizes the concept of context-space. The DRL-RM included two RL agents: 1) computing policy agent that applied actor-critic algorithm (DDPG); 2) radio policy agent that deployed unsupervised deep neural network with multiple input neurons (i.e., encoded data, the output of computing policy, and feedback from the output) and single output neuron (i.e., radio control decision). DRL-RM's outputs were utilized to configure computing and radio schedulers. DRL-RM managed both schedulers by allocating computing and radio resources according to the wireless environment's current conditions and computing resource availability. 

The problem of diversity in resources for RAN slicing design was addressed in~\cite{EdgeSliceWithDDRL} by the deployment of resource autonomy (RA) agent at each slice. The authors synthesized a performance coordination (PC) function at the mobile edge computing (MEC) to orchestrate the distributed RAs. They modeled PC using a stochastic programming tool, while RAs' DRL agents deployed the DDPG algorithm.

Both works in~\cite{DL_based_Comp_RAN_Scheduling_Mobicom19, EdgeSliceWithDDRL} adopted the concept of multi-agent RL but in different flavors. The former introduced it in the context of diversity in resources (radio and computing), while the latter discussed it in the light of distributed (decentralized) agents. However, the bottleneck in transport resources (fronthaul and backhaul links) were not examined. Additionally, authors in~\cite{DL_based_Comp_RAN_Scheduling_Mobicom19} abstracted computing resources only in CPU cycles; however, the work in~\cite{EdgeSliceWithDDRL} considered storage resources. Nevertheless, both did not address the allocation of caching resources.
Furthermore,~\cite{DL_based_Comp_RAN_Scheduling_Mobicom19} did not discuss the necessity of using multiple slice controllers, as discussed earlier in ``The Concept of Network Slicing.'' Instead of developing a slice controller placement method to address the distribution of resources,  authors in~\cite{EdgeSliceWithDDRL} introduced a distributed slicing scheme (RA agent per slice) that needs coordination with a centralized performance controller (PC). The challenge is the lack of reallocation periodicity analysis in the PC optimization, where its auxiliary and dual variables relied on released information from RAs. Thus, the coordination time between PC and RAs should be studied for real-time resource (re)-allocation.   

Authors in~\cite{RL4DRA_RAN_Slicing} addressed the dynamics in the wireless environment by introducing an RL-based spectrum sharing scheme for RAN-S. Furthermore, in~\cite{Intelligent_Rsrc_Schedu_5G_RAN_Slicing}, authors recalled the significance of RU prediction (RUP) for RAN-S. They incorporated DL-based RUP into an RL-based scheduling algorithm to address the dynamics in RAN environment in long and short time scales. The scheduling problem was formulated as a continuous Markovian Decision Process (MDP), and the asynchronous advantage actor-critic (A3C) algorithm was used for agent learning. Each slice deployed A3C on a parallel computing platform to schedule spectrum resources. Nevertheless, both works in~\cite{RL4DRA_RAN_Slicing, Intelligent_Rsrc_Schedu_5G_RAN_Slicing} did not address the resources' heterogeneity and the action-space's continuity.

Accordingly, to address the dynamics in resources for iRAN-S, we envisage that DRA would adopt a multi-agent learning framework, e.g., multi-agent RL (MARL), as shown in Figure~\ref{fig:DRAandSchedu}. However, as we discuss later, further research is still needed to address its technical challenges.

\subsection*{Slice Isolation}

\begin{figure}
 \begin{center}
 \includegraphics[width=1\columnwidth]{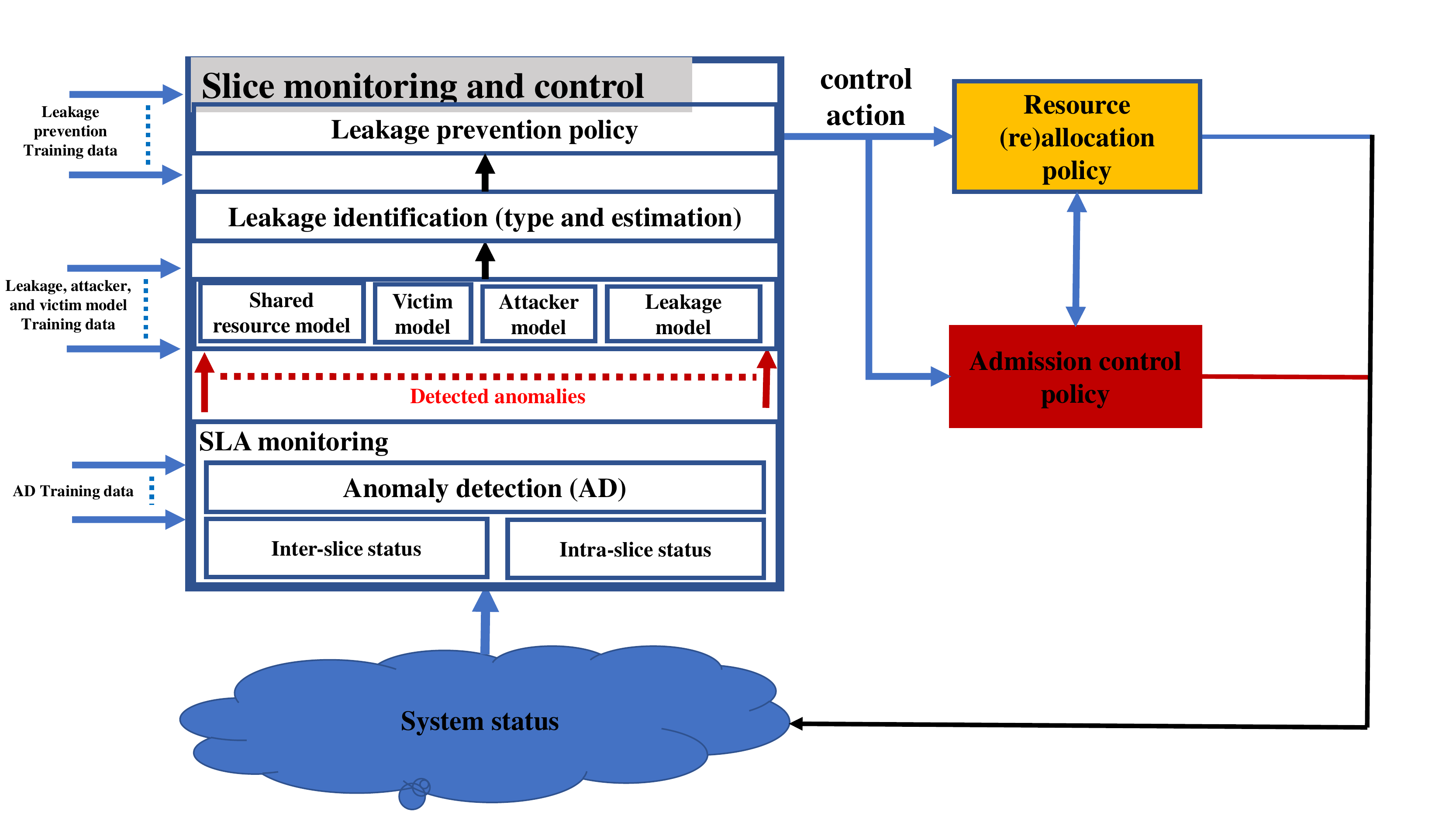}
 \caption{Slice isolation from information leakage perspective~\cite{NN_autoencoder_Trans_Info_Forensics_Security, SCAUL}.}
 \label{fig:Sisolate}
\end{center}
\end{figure}

As explained earlier, \textit{slice isolation} aims to maintain the tenant's SLA.  Securing slices and preserving users' privacy are significant components of SLA. In the following, we discuss slice isolation from the security design perspective.

A shared environment would facilitate establishing side-channels across slices leading to side-channel attacks (SCAs). To exemplify, consider two slices with two VNFs instances that run on the same resources. A malicious VNF in one slice may extract fine-grained information, ``information leakage (IL),'' from a victim VNF in the other slice. Thus, SCA is defined by modeling IL, attacker, and client (victim) behavior.

RAN slicing would be prone to two SCAs categories: 1) wireless communication SCA; 2) computing SCA. While channel coding schemes would scrutinize the former, e.g.,~\cite{NN_autoencoder_Trans_Info_Forensics_Security}, the latter constitutes diverse SCA models that are defined based on how the information leaks from shared resources. Different SCA models were discussed in the context of shared computing resources, such as cloud computing and system-on-chip (SoC). SCA models include timing, caching, co-residency, power, electromagnetic, acoustic, differential-fault cryptanalysis, data-remanence, and optical SCA models.

Figure~\ref{fig:Sisolate} envisages a generic slice monitoring and control agent in a slice controller for iRAN-S. It would include SLA monitoring to detect any anomalies in inter-slice and intra-slice status. According to detected anomalies, the slice security agent would utilize trained agents (shared resource, victim, attacker, and IL models) for IL detection and estimation. Accordingly, the IL agent would apply a leakage prevention policy to determine the best action for slice protection and update the resource (re)allocation and AC agents.

Since SCAs hinge on IL, we review the state-of-the-art ML-based IL models of two different SCA models (wireless channel SCA and computing SCA).

To provide a flexible channel coding that would accomplish reliable and IL-free wireless transmission, an FFNN has been introduced in~\cite{NN_autoencoder_Trans_Info_Forensics_Security}. It synthesized an intelligent encoder and decoder to maximize the reliability, i.e., minimize the block length error rate (BLER) and minimize IL. The problem was formulated as an unconstrained weighted sum multi-optimization problem. Additive-White-Gaussian-Noise (AWGN) modeled the wireless channel (i.e., shared resource model). The Monte Carlo simulation prediction phase was utilized to provide several realizations to estimate BLER at the receiver end. The authors estimated IL based on second-order Taylor expansion of the received Gaussian Mixture's differential entropy at the adversary side. Accordingly, the FFNN encoder got a message and a uniformly random bit confusion message at its input. It returned estimated bits at the legitimate receiver and the code word that went through the AWGN channel for message transmission. They also used the FFNN encoder for IL estimation.

Authors in~\cite{SCAUL} developed a multi-layer-perceptron (MLP) NN-based sensitivity analysis to derive the leakage model. The observed traces (measured leaked information) were clustered, then encoded using long-short-term memory (LSTM) autoencoder to provide contextual features (secret keys patterns). The MLP-based sensitivity analysis used the patterns to derive a leakage model for every secret keys candidate. Then, MLP's weights were mildly perturbed, which led to alterations of its output. Data features with the lowest variance included the leakage model. Subsequently, the secret keys patterns were clustered based on the derived leakage model for each candidate key. The candidate key that returned the highest Hamming distance (i.e., inter-cluster distance) was selected as the correct secret key. The leakage model for this system was identified after obtaining the right key. Thus, a further attacker (threat) and client modeling procedures would be conducted to provide SCA countermeasures.

Table~\ref{tab:summary} summarizes the opportunities and the associated challenges of incorporating ML into RAN-S.


\begin{table*}[ht]
\centering
\begin{tabular}{c}
\includegraphics[width=2\columnwidth]{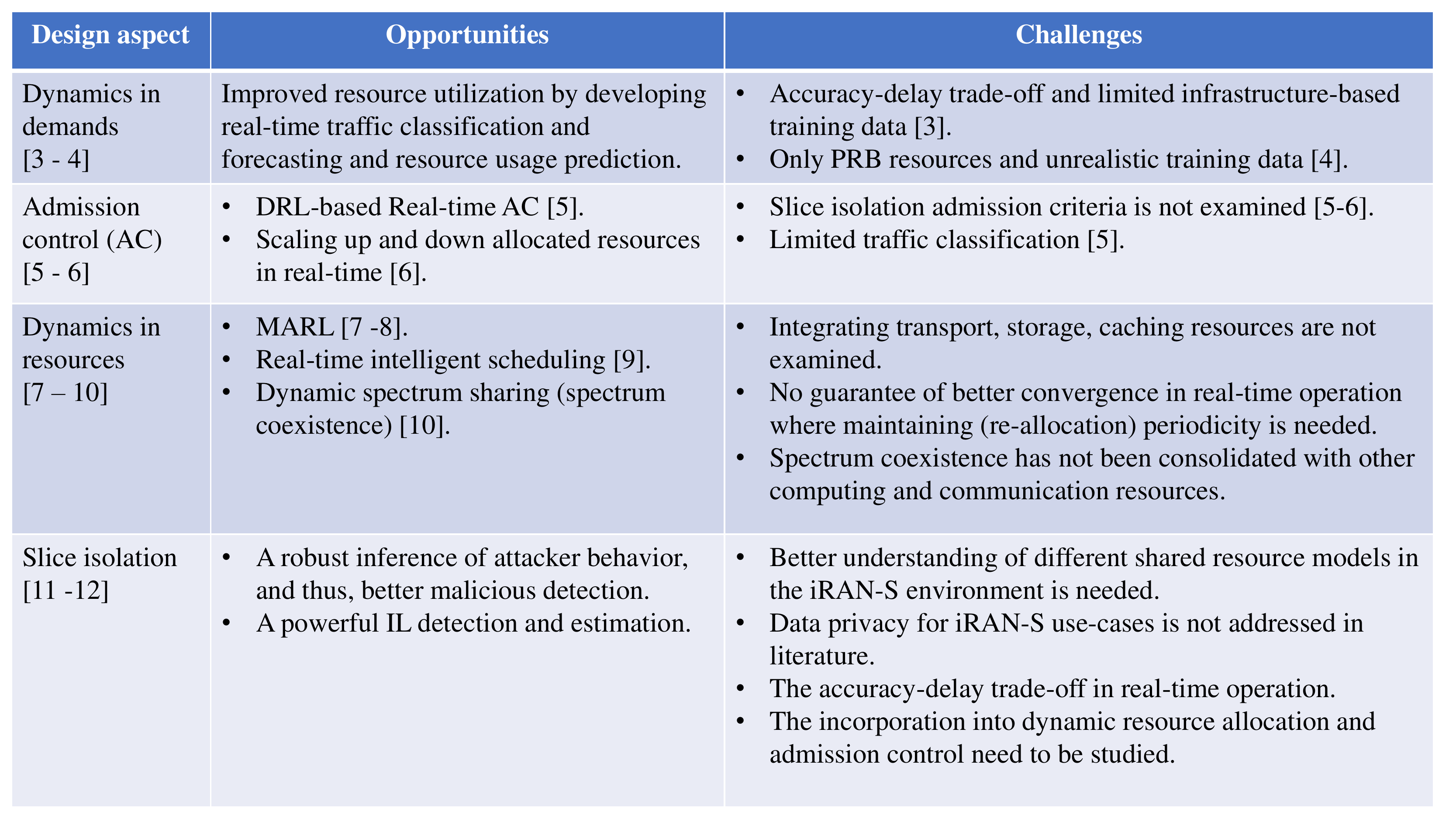}
\end{tabular}
\caption{ML's opportunities and challenges for iRAN-S}
\label{tab:summary}
\end{table*}


\section*{Directions for future research}
In the following, we discuss a few directions for future research towards iRAN-S.

\textbf{Virtualization:} As explained earlier, virtualization relies on the software environment in iRAN-S, which embraces SDN and SDR. Although SDN runs on general-purpose-processing (GPP) units, SDR runs on different architectures for fast signal processing of different waveform applications, i.e., SDR runs on Field-programmable-gate-array (FPGA) and GPP architectures. 
Running various signal processing functions (e.g., different waveform shaping techniques from other running slices) on shared heterogeneous SDR architectures brings the challenge of irregular runtime adaptation (IRA) to the iRAN-S scene. IRA was examined for graph algorithms design in high-performance computing (HPC)~\cite{IRE_Atharva}. In~\cite{IRE_Atharva}, authors discussed the runtime adaptation between various parallelizing techniques for irregular applications that run across shared heterogeneous architectures. Irregularity at runtime could arise due to several reasons, such as irregular memory access patterns (IMA) and irregular workload distribution (IWD). In IMA, it is unknown which processor unit will access data because it is only determined at runtime.
Likewise, the memory locations, from where to retrieve the data, are also determined at runtime. In IWD, the application's nature is such that we can not distribute the workload evenly across multiple heterogeneous shared cores. In the virtualized wireless computing domain, radio signal processing functions and the interpretation of signals are different from typical data processing. Thus, we need to revisit the IRA for wireless virtualization in iRAN-S.

\textbf{Data privacy:} Federated learning (FL) would augment the DRL-based DRA for distributed data training. FL-based local training would be utilized to transmit data parameters (e.g., gradients) to DRL agents instead of transmitting real-data that would cause bandwidth and delay overheads. However, there are some cases that machines might have limited processing power, and thus, data transmission becomes critically needed. This scenario brings the privacy challenge to the scene. Motivated by the remarkable success of differential privacy (DP)~\cite{Diff_Privacy} in protecting sensitive personal data, DP could be introduced for iRAN-S. Different DP settings deem applicable to iRAN-S, such as slicing-based smart home use-case, in which collecting aggregated statistics from smart home meters is needed. In this setup, a collection of devices jointly contribute to training an FL model without disclosing the data locally accessible by each device. Hence, DP perturbation mechanisms could be employed for inference and analysis while preserving data privacy. However, this hypothesis is still in its infancy and requires further research.

\textbf{Slice isolation and customization trade-off:} iRAN-S aims to provide a rapid and flexible reconfiguration of network settings for virtual RAN slices to fulfill various SLAs (i.e., slice customization). Nevertheless, there is a trade-off between allowing prompt instantiation and guaranteeing slice isolation. As discussed earlier, the required time to instantiate a slice to meet SLA mainly hinges on heterogeneous resources' scheduling time and the AC algorithm's complexity. On the other hand, the timing SCA model is present in shared scheduler environments as in iRAN-S. Hence, iRAN-S opens another direction for researching DRA-based timing IL detection and prevention. This research would aim to mitigate timing SCA's impact on slice isolation performance and concurrently attain prompt slice customization.

\textbf{Multi-agent Reinforcement Learning (MARL):} MARL would be the best candidate for iRAN-S because of its capability of managing various control environments, i.e., different wireless technologies, distributed edge computing, and centralized control (SDN). Nevertheless, MARL introduces three design challenges that need further research: 1) communication protocol between agents; 2) novel algorithms that would handle the discrepancies in time requirements for allocating different resources (i.e., sub-milliseconds for radio resources and few seconds for computing resources); 3) developing synchronized learning algorithms~\cite{MARL_ICLR20_Sandeep}.

\section*{Conclusion}
This article reviews the recent advances in machine learning techniques that endowed opportunities towards intelligent radio access network (RAN) slicing design (iRAN-S), such as traffic forecasting, admission control, dynamic resource allocation (DRA), scheduling, and information leakage models for slice isolation. We outline critical design challenges for each introduced machine learning technique from theoretical and practical perspectives. We conclude that virtualization, the privacy of training data, and the isolation-customization trade-off open further research directions. Besides, we emphasize that multi-agent reinforcement learning would be adopted for iRAN-S; however, further research is needed for agents' communication and learning algorithms.


\bibliographystyle{IEEEtran}
\bibliography{IEEEabrv,Towards_Intelligent_RAN_Slicing_for_B5G_Opportunities_and_Challenges}

\begin{thebibliography}{10}
\providecommand{\url}[1]{#1}
\csname url@rmstyle\endcsname
\providecommand{\newblock}{\relax}
\providecommand{\bibinfo}[2]{#2}
\providecommand\BIBentrySTDinterwordspacing{\spaceskip=0pt\relax}
\providecommand\BIBentryALTinterwordstretchfactor{4}
\providecommand\BIBentryALTinterwordspacing{\spaceskip=\fontdimen2\font plus
\BIBentryALTinterwordstretchfactor\fontdimen3\font minus
  \fontdimen4\font\relax}
\providecommand\BIBforeignlanguage[2]{{%
\expandafter\ifx\csname l@#1\endcsname\relax
\typeout{** WARNING: IEEEtran.bst: No hyphenation pattern has been}%
\typeout{** loaded for the language `#1'. Using the pattern for}%
\typeout{** the default language instead.}%
\else
\language=\csname l@#1\endcsname
\fi
#2}}
\renewcommand\BIBentryALTinterwordstretchfactor{4}

\bibitem{CompInComm}
F.~{Fitzek}, F.~{Granelli}, and P.~{Seeling}, \emph{"Computing in Communication
  Networks: From Theory to Practice"}.\hskip 1em plus 0.5em minus 0.4em\relax
  Amsterdam: Academic Press, 2020.

\bibitem{NS_meets_AI}
D.~{Bega} \emph{et~al.}, ``{Network Slicing Meets Artificial Intelligence: An
  AI-Based Framework for Slice Management},'' \emph{IEEE Commun. Mag.},
  vol.~58, no.~6, pp. 32--38, 2020.

\bibitem{DeepAI4MTDMobicom20}
C.~{Zhang} \emph{et~al.}, ``Microscope: {Mobile Service Traffic Decomposition
  for Network Slicing as a Service},'' in \emph{Proc. ACM MobiCom}, Sep 2020,
  pp. 503 -- 516.

\bibitem{RsrcUsagePredictionMobiHoc19}
{C. Gutterman} \emph{et~al.}, ``{RAN Resource Usage Prediction for a 5G Slice
  Broker},'' in \emph{Proc. ACM Mobihoc}, Jul 2019, p. 231–240.

\bibitem{ML_Based_AC_5G_MarketOptim}
D.~{Bega} \emph{et~al.}, ``{A Machine Learning Approach to 5G Infrastructure
  Market Optimization},'' \emph{IEEE Trans. Mobile Computing}, vol.~19, no.~3,
  pp. 498--512, 2020.

\bibitem{RL_Flexible_AC}
M.~R. {Raza} \emph{et~al.}, ``{Reinforcement Learning for Slicing in a 5G
  Flexible RAN},'' \emph{Journal of Lightwave Technology}, vol.~37, no.~20, pp.
  5161--5169, 2019.

\bibitem{DL_based_Comp_RAN_Scheduling_Mobicom19}
J.~A. Ayala-Romero \emph{et~al.}, ``{VrAIn: A Deep Learning Approach Tailoring
  Computing and Radio Resources in Virtualized RANs},'' in \emph{Proc. ACM
  MobiCom}, Oct. 2019, pp. 1 -- 16.

\bibitem{EdgeSliceWithDDRL}
Q.~Liu, T.~Han, and E.~Moges, ``{EdgeSlice: Slicing Wireless Edge Computing
  Network with Decentralized Deep Reinforcement Learning},'' 2020.

\bibitem{RL4DRA_RAN_Slicing}
Y.~{Shi}, Y.~E. {Sagduyu}, and T.~{Erpek}, ``{Reinforcement Learning for
  Dynamic Resource Optimization in 5G Radio Access Network Slicing},'' in
  \emph{Proc. IEEE CAMAD}, Sept. 2020, pp. 1--6.

\bibitem{Intelligent_Rsrc_Schedu_5G_RAN_Slicing}
M.~{Yan} \emph{et~al.}, ``{Intelligent Resource Scheduling for 5G Radio Access
  Network Slicing},'' \emph{IEEE Trans. Vehicular Technology}, vol.~68, no.~8,
  pp. 7691--7703, 2019.

\bibitem{NN_autoencoder_Trans_Info_Forensics_Security}
K.~{Besser} \emph{et~al.}, ``{Wiretap Code Design by Neural Network
  Autoencoders},'' \emph{IEEE Trans. on Information Forensics and Security},
  vol.~15, pp. 3374--3386, 2020.

\bibitem{SCAUL}
{K. Ramezanpour}, {P. Ampadu}, and {W. Diehl}, ``{SCAUL: Power Side-Channel
  Analysis With Unsupervised Learning},'' \emph{IEEE Trans. on Computers},
  vol.~69, no.~11, pp. 1626--1638, Nov. 2020.

\bibitem{IRE_Atharva}
A.~{Gondhalekar} and W.~C. {Feng}, ``{Exploring FPGA Optimizations in OpenCL
  for Breadth-First Search on Sparse Graph Datasets},'' in \emph{Proc. Intl
  Conf. FPL}, 2020, pp. 133--137.

\bibitem{Diff_Privacy}
T.~{Zhang} \emph{et~al.}, ``{Correlated Differential Privacy: Feature Selection
  in Machine Learning},'' \emph{IEEE Trans. on Industrial Informatics},
  vol.~16, no.~3, pp. 2115--2124, 2020.

\bibitem{MARL_ICLR20_Sandeep}
T.~{Chu}, S.~{Chinchali}, and S.~{Katti}, ``{Multi-agent Reinforcement Learning
  for Networked System Control},'' in \emph{Proc. ICLR}, 2020, pp. 1--9.

\end{thebibliography}

\vspace{0.2cm}
\noindent
\textbf{EmadElDin A. Mazied} is a Ph.D. student in the Bradley Department of Electrical and Computer Engineering at Virginia Tech, USA. He is an assistant lecturer (on-leave) in the Electrical Engineering Department, Sohag University, Egypt. His research interests include resource allocation, stochastic optimization, machine learning, and software-defined networks.

\vspace{0.2cm}
\noindent
\textbf{Lingjia Liu} is an Associate Professor in the Bradley Department of Electrical and Computer Engineering and is serving as the Associate Director of Wireless@VT at Virginia Tech, USA. His research interests include machine learning for wireless communications, enabling technologies for 5G and beyond, mobile edge computing, and Internet of Things.

\vspace{0.2cm}
\noindent
\textbf{Scott F. Midkiff} is a professor and vice president for information technology \& chief information officer at Virginia Tech. He was electrical and computer engineering department head (2009-2012) at Virginia Tech and a program director at the National Science Foundation (2006-2009). His research interests include wireless networks and cyber-physical systems.
\end{document}